%% file: main.tex
\documentclass[runningheads]{llncs}
\usepackage[dvipsnames]{xcolor}
\usepackage{graphicx}
\usepackage{amssymb}
\usepackage[english]{babel}
\usepackage{todonotes}
\usepackage{hyperref}
\usepackage{placeins}
\usepackage{url}
\usepackage{multirow}
\usepackage{verbatim}
\usepackage{fancyvrb}
\usepackage{xspace}
\usepackage{booktabs}
\usepackage{graphicx}
\usepackage{eurosym}
\usepackage{inconsolata}
\usepackage{float}
\usepackage{amsmath}
\usepackage{breqn}
\usepackage{wrapfig,lipsum,booktabs}

\newcommand{\greenbrain}{\xspace{\sf\small \textcolor{PineGreen}{\textbf{G}}FogBrain}\xspace}
\newcommand{\cd}[1]{\texttt{\small #1}\xspace}
\newcommand{\co}{$\text{CO}_2$\xspace}

\newcommand{\kgcokw}{$\text{kgCO}_2\text{/kWh}$\xspace}
\newcommand{\kgco}{$\text{kgCO}_2$\xspace}
\begin{document}
\title{Green Application Placement\\ in the Cloud-IoT Continuum
}
%
%
\author{Stefano Forti\orcidID{0000-0002-4159-8761} \and
Antonio Brogi\orcidID{0000-0003-2048-2468}}
\authorrunning{S. Forti and A. Brogi}
%
\institute{Department of Computer Science, University of Pisa, Italy\\
\email{\{stefano.forti, antonio.brogi\}@di.unipi.it}\\}
\maketitle              
\vspace{-5mm}
\begin{abstract}
Green software engineering aims at reducing the environmental impact due to developing, deploying, and managing software systems. Meanwhile, Cloud-IoT paradigms can contribute to improving energy and carbon efficiency of application deployments by (\textit{i}) reducing the amount of data and the distance they must travel across the network, (\textit{ii}) by exploiting idle edge devices to support application deployment. In this article, we propose a declarative methodology and its Prolog prototype for determining placements of application services onto Cloud-IoT infrastructures so to optimise energy and carbon efficiency, also considering different infrastructure power sources and operational costs. The proposal is assessed over a motivating example.
\end{abstract}
\input{src/introduction.tex}
\input{src/methodology.tex}
\input{src/relatedwork.tex}

\input{src/conclusions.tex}

\newpage
%
%
%
\bibliographystyle{splncs04}
\bibliography{biblio}
\newpage
\input{src/appendix.tex}

\end{document}

%% file: src/introduction.tex
\vspace{-8mm}
\section{Introduction}
\label{sec:introduction}
\vspace{-2mm}
ICT energy demand could possibly reach 14\% of the total worldwide footprint by 2040 \cite{belkhir2018assessing}. As climate scientists agree on the urgency of reducing the human impact on the environment, \textit{green software engineering} is getting increasing attention as a possible way to contain ICT energy usage and carbon emissions, through achieving a more sustainable software life-cycle~\cite{calero2015green}. While much work has focussed on embedding sustainability principles in software design phases, less work has proposed methodologies and tools to improve and assess software lifecycle sustainability~\cite{greensweng}, i.e. from application testing to deployment and runtime management.
Meanwhile, Cloud-IoT computing paradigms -- e.g. Fog, Edge computing \cite{brogi2018bonsai} -- have been proposed to improve on the Quality of Service (QoS) of emerging latency-sensitive and bandwidth-hungry applications. As highlighted by some authors \cite{sarkar2016theoretical,kopras2020latency}, those paradigms can also represent greener alternatives to the Cloud paradigm as they can exploit pervasive and possibly idle computational devices closer to the IoT, thus improving on energy efficiency of those idle resources and reducing unnecessary data transfer from/to the Cloud.

To achieve the above sustainability goals, it is crucial to place application services so to meet all their requirements \textit{and} by determining the best trade-off between the operation costs of their deployment and the expected energy consumption and carbon emissions, which very much depend on the characteristics of the target deployment nodes (i.e. energy profile, power sources, power usage effectiveness). While the problem of placing application services onto Cloud-IoT infrastructure to meet their software, hardware, IoT and network requirements has been extensively studied \cite{applicationmanagementfogsurvey,brogi2019place}, the problem of determining energy- and carbon-aware placements was only marginally addressed until very recently \cite{aldossary2021towards}.

In this article, based on our previous work \cite{fogbrain}, we illustrate a declarative programming solution to the problem of determining energy- and carbon-aware application placements in Cloud-IoT settings, also capable of estimating operational deployment costs. Our methodology permits determining application placements that meet software, hardware, IoT, latency and bandwidth requirements, and to estimate deployment costs, energy consumption and carbon emissions, relying on data disclosed on the available Cloud-IoT nodes. Being declarative, our approach is easy to understand and to extend, e.g. by employing alternative formulas to estimate all of the above. A Prolog open-source prototype, \greenbrain, is assessed over a motivating example based on lifelike data. 

The rest of this article is organised as follows. Section \ref{sec:methodology} describes the model and methodology of \greenbrain, while showcasing its functioning over a lifelike motivating example. Section \ref{sec:related} briefly discusses some closely related work, and Section \ref{sec:conclusions} concludes by pointing to some directions for future work.

%% file: src/methodology.tex
\vspace{-3mm}
\section{\xspace{\sf \textcolor{PineGreen}{\textbf{G}}FogBrain}\xspace in Action}
\label{sec:methodology}

In this section, we illustrate \greenbrain's prototype and methodology by means of a lifelike motivating example from \textit{smart environments} \cite{9556307}.
\greenbrain extends our previous work in the field of context- and QoS-aware placements of Cloud-IoT applications, to also determine energy- and carbon-aware placements. Particularly, we extend the model prototype of~\cite{fogbrain} to consider all necessary ingredients to estimate energy consumption and carbon emissions of running applications, and operational costs (i.e. due to leasing computational resources to keep application services up and running). 

Our goal is to support application operators, enabling them to informedly identify placements that can reduce energy consumption and carbon emissions, while assessing the impact that \textit{being greener} could have on the operational costs of their deployments.
In the next paragraphs, we detail our declarative application and infrastructure model and the declarative programming methodology implemented by \greenbrain to achieve such a goal.

\vspace{-3mm}
\noindent
\paragraph{\textbf{Applications Requirements} --} As in \cite{fogbrain}, application \cd{A} made of services \cd{S1} \dots \cd{Sk} is declared as

\begin{Verbatim}[fontfamily=zi4, fontsize=\footnotesize, frame=single, framesep=1mm, framerule=0.1pt, rulecolor=\color{gray}, tabsize=4]
application(A, [S1, ..., Sk]).
\end{Verbatim}

\noindent
The software, hardware\footnote{For the sake of simplicity, we represent hardware units as integers as in \cite{fogbrain}.} and IoT requirements of service \cd{S} are declared as

\begin{Verbatim}[fontfamily=zi4, fontsize=\footnotesize, frame=single, framesep=1mm, framerule=0.1pt, rulecolor=\color{gray}, tabsize=4]
service(S, SoftwareReqs, HardwareReqs, IoTReqs).
\end{Verbatim}

\noindent Finally, interactions between services \cd{S1} and \cd{S2} with associated maximum end-to-end latency and minimum bandwidth requirements are declared as

\begin{Verbatim}[fontfamily=zi4, fontsize=\footnotesize, frame=single, framesep=1mm, framerule=0.1pt, rulecolor=\color{gray}, tabsize=4]
s2s(S1, S2, MaxLatency, MinBandwidth).
\end{Verbatim}

\begin{figure}
    \centering
    \includegraphics[width=0.7\textwidth]{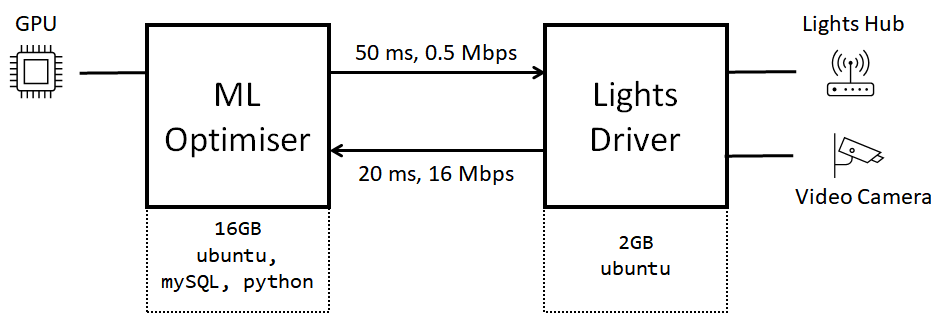}
    \caption{Example application.}
    \label{fig:app}
\end{figure}

\example\label{ex:app} The application of Fig. \ref{fig:app} consists of two interacting services -- \textsf{Lights Driver} and \textsf{ML Optimiser} -- for optimising ambient lighting in a museum based on processing real-time video footage. 
The \textsf{Lights Driver} requires 2GB and Ubuntu to run, and to reach out a video-camera and a lights hub. Similarly, \textsf{ML Optimiser} requires 16GB of RAM and the availability of Ubuntu, MySQL and Python on the deployment node, which must also be equipped with a GPU for processing streamed data. 
Finally, the interaction from the \textsf{Lights Driver} to the \textsf{ML Optimiser} require at least 16Mbps of available bandwidth and tolerates at most 20 ms latency. On the other hand, the the interaction from the  \textsf{ML Optimiser} to the \textsf{Lights Driver} needs only 0.5 Mbps, with a latency lower than 50 ms.
Such an application can be simply declared as in Fig. \ref{fig:appcode}. \hfil\qed

\vspace{-5mm}

\vspace{-13mm}
\begin{figure}
    \centering
\begin{Verbatim}[fontfamily=zi4, fontsize=\footnotesize, frame=single, framesep=1mm, framerule=0.1pt, rulecolor=\color{gray}, tabsize=4]
application(lightsApp, [mlOptimiser, lightsDriver]).
    service(mlOptimiser, [mySQL, python, ubuntu], 16, [gpu]).
    service(lightsDriver, [ubuntu], 2, [videocamera, lightshub]).
    s2s(mlOptimiser, lightsDriver, 50, 0.5).
    s2s(lightsDriver, mlOptimiser, 20, 16).
\end{Verbatim}
\vspace{-4mm}
    \caption{Example application declaration.}
    \label{fig:appcode}
\end{figure}

\vspace{-10mm}
\noindent
\paragraph{\textbf{Infrastructure capabilities} --} Complementarily to application service requirements, Cloud-IoT nodes can be declared along with their software, free hardware and IoT capabilities, and with the unit hourly cost for leasing hardware resources:

\begin{Verbatim}[fontfamily=zi4, fontsize=\footnotesize, frame=single, framesep=1mm, framerule=0.1pt, rulecolor=\color{gray}, tabsize=4]
node(NodeId, SoftwareCapabilities, FreeHW, IoTCapabilities).
cost(NodeId, UnitHWCostPerHour).
\end{Verbatim}

\noindent
Similarly, end-to-end links between nodes \cd{N1} and \cd{N2} are declared, along with their \cd{FeaturedLatency} and \cd{FeaturedBandwidth}, as in

\begin{Verbatim}[fontfamily=zi4, fontsize=\footnotesize, frame=single, framesep=1mm, framerule=0.1pt, rulecolor=\color{gray}, tabsize=4]
link(N1, N2, FeaturedLatency, FeaturedBandwidth).
\end{Verbatim}

The \textit{power usage effectiveness} (PUE) associated to a node is the ratio between the overall energy needed for keeping the node working and the energy that the node uses for actual computation. For instance, a PUE of 1.5 indicates that for every 1kWh spent in computation, another 0.5kWh is needed for non-IT tasks (e.g. cooling, lighting, network) that keep the server working. Typical values of the PUE range between $1.2$ and $1.9$. 
Extending the model of \cite{fogbrain}, we assume that node operators can disclose information about the total hardware (free and in use) at each node and the associated PUE as in 

\begin{Verbatim}[fontfamily=zi4, fontsize=\footnotesize, frame=single, framesep=1mm, framerule=0.1pt, rulecolor=\color{gray}, tabsize=4]
totHW(N,TotalHardware).                 pue(N,PUE).
\end{Verbatim}

\noindent

Node operators can then specify the energy consumption profile of each node via predicates like

\begin{Verbatim}[fontfamily=zi4, fontsize=\footnotesize, frame=single, framesep=1mm, framerule=0.1pt, rulecolor=\color{gray}, tabsize=4]
energyProfile(N,Load,EnergyConsumption) :- ...
\end{Verbatim}

\noindent where \cd{EnergyConsumption} is obtained in kWh as a, possibly non-linear, function of the current percentage \cd{Load} at node \cd{N}. Existing processors show a baseline energy consumption even when they are idle, which increases as the node workload increases \cite{microsoft}. 

Last, the percentages of the energy mix of each node can be specified as in

\begin{Verbatim}[fontfamily=zi4, fontsize=\footnotesize, frame=single, framesep=1mm, framerule=0.1pt, rulecolor=\color{gray}, tabsize=4]
energySourceMix(N,[(P1,Source1), ..., (PK,SourceK)]).
\end{Verbatim}

\vspace{-3mm}
\noindent where \cd{PJ} is the percentage of electricity that node \cd{N} receives from \cd{SourceJ}. 

We finally assume that average \co emissions for each source are declared in a public knowledge base of facts like \cd{emissions(Source, Mu)}, where \cd{Mu} are the emissions in \kgcokw for \cd{Source}, e.g. as those reported in Table \ref{tab:emissions}. 

Note that, when energy-related information is not disclosed, \greenbrain easily allows to employ default data or data taken from public audits such as \cite{greenpeace}. 

\vspace{-5mm}
\begin{table}
\centering
    \caption{\co emissions per power source \cite{cotwo}.}
    \label{tab:emissions}
{\footnotesize
\begin{tabular}{|c|c|}
\hline
\multicolumn{1}{|c|}{\textbf{Power Source}} & \multicolumn{1}{c|}{\textbf{Emissions [$\text{kgCO}_2\text{/kWh}$]}} \\ \hline
\textit{gas}                                & 0.610                                                                \\ \hline
\textit{coal}                               & 1.100                                                                \\ \hline
\textit{on shore wind}                      & 0.0097                                                               \\ \hline
\textit{off shore wind}                     & 0.0165                                                               \\ \hline
\textit{solar}                              & 0.05                                                                 \\ \hline
\end{tabular}}
\end{table}

\begin{figure}[!h]
    \centering
    \includegraphics[width=0.7\textwidth]{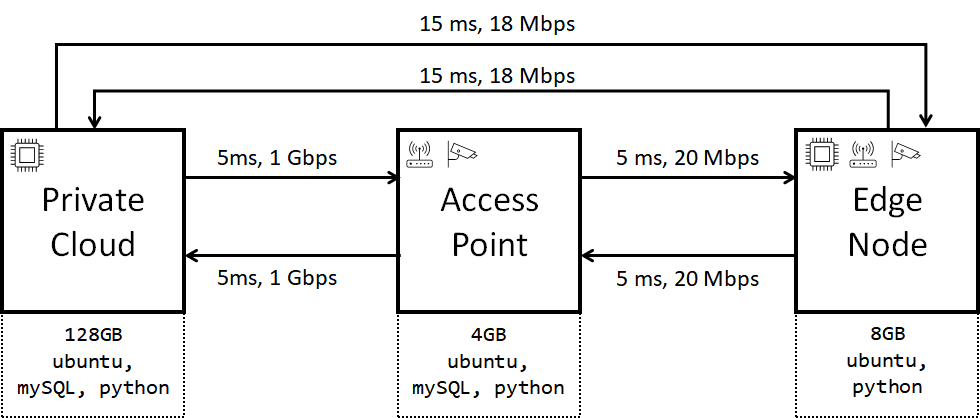}
    \caption{Example Cloud-IoT infrastructure.}
    \label{fig:infra}
\end{figure}

\vspace{-6mm}
\example\label{ex:infra} Consider the Cloud-IoT infrastructure of Fig. \ref{fig:infra} to deploy the application of Example \ref{ex:app}. Fig. \ref{fig:infracode} epitomises the declaration of the capabilities and energy information of all three nodes. We only show the declaration of the link between \textsf{Private Cloud} and \textsf{Access Point}.

\begin{figure}
    \centering
\begin{Verbatim}[fontfamily=zi4, fontsize=\footnotesize, frame=single, framesep=1mm, framerule=0.1pt, rulecolor=\color{gray}, tabsize=4]
node(privateCloud,[ubuntu, mySQL, python], 128, [gpu]).
    cost(privateCloud,0.0016).      totHW(privateCloud,150).
    energyProfile(privateCloud,L,E) :- E is 0.1 + 0.01*log(L).
    pue(privateCloud,1.9).
    energySourceMix(privateCloud,[(0.3,solar), (0.7,coal)]).
node(accesspoint,[ubuntu, mySQL, python], 4, [lightshub, videocamera]).
    cost(accesspoint,0.003).        totHW(accesspoint,6).
    energyProfile(accesspoint,L,E) :- E is 0.05 + 0.03*log(L).
    pue(accesspoint,1.5).
    energySourceMix(accesspoint,[(0.1,gas),(0.8,coal),(0.1,onshorewind)]).
node(edgenode,[ubuntu, python], 8, [gpu, lightshub, videocamera]).
    cost(edgenode,0.005).       totHW(edgenode,12).
    energyProfile(edgenode,L,E) :- L =< 50 -> E is 0.08; E is 0.1.
    pue(edgenode,1.2).
    energySourceMix(edgenode,[(0.5,coal), (0.5,solar)]).
 
link(privateCloud, accesspoint, 5, 1000).
\end{Verbatim}
\vspace{-4mm}
    \caption{Example infrastructure declaration.}
    \label{fig:infracode}
\end{figure}

Note that, for instance, node \textsf{Private Cloud} currently features 128 free hardware units (out of the 150 totally available), each offered for 0.0016 cents per hour, and that its energy consumption in kWh is given by a function $\phi(L)$ of the current workload $L$ such that $\phi(w) = 0.1 \text{kWh} + 0.01 \cdot \log(L)$ with $L \in [0,100]$. Besides, \textsf{Private Cloud} is powered by an energy mix coming for $30\%$ from a solar plant and for $70\%$ from a coal plant, as declared by \cd{energySourceMix/2}. Last, the PUE of \textsf{Private Cloud} is 1.9.
\hfill\qed

\paragraph{\textbf{Energy-, Carbon- and Cost-aware Placements} --} Fig. \ref{fig:gFogBrain} shows\footnote{Due to space limitations, we only show the main predicates of \greenbrain. Full code is open-sourced at \url{https://github.com/di-unipi-socc/fogbrainx/tree/main/green}.} how the \greenbrain prototype determines energy- and carbon-aware application placements in Cloud-IoT settings. Predicate \cd{placements/2} (lines 1--3) determines all \cd{Placements} that satisfy software, hardware, IoT and network QoS requirements of the application by means of \cd{gFogBrain/4}, along with the associated hourly deployment \cd{Cost}, energy consumption \cd{E} and carbon emissions \cd{C} (line 2). 
The obtained placements are sorted by increasing estimated carbon footprint, cost, and energy consumption, considered in this order of priority\footnote{By suitably rearranging output tuples, it is possible to prioritise differently among the estimated metrics. For instance, the order \cd{(Cost,E,C,P)} at line 2 would give priority to cost over energy consumption and carbon emissions.} (line 3).

Predicate \cd{gFogBrain/4} (lines 4--6) exploits the \cd{placement/2} (line 5) and the \cd{allocatedResources/2} (line 6) predicates of \cite{fogbrain} (see Appendix A) to determine a placement that satisfies software, hardware, IoT and network QoS requirements of a given application, and the associated hardware and bandwidth in use, respectively. Then, \cd{Energy} consumption and \cd{Carbon} emissions associated to the placement are computed via the \cd{footprint/4} predicate (lines 6, 7--11).

Based on the deployment nodes used by \cd{P} (line 8), the predicate \cd{footprint/4} computes hardware- (line 9) and network-related (line 10) energy consumption and carbon emissions and sums them, respectively (line 11).
\greenbrain employs an extended version of the model from Kelly et al. \cite{cotwo} to associate an estimate of energy consumption to a piece of computation running on a given node. The energy consumption $\cd{E}_{s}$ of a service $s$ running on a node $n$ is computed as

\begin{equation}\label{eq:energy}
\cd{E}_{s} = \cd{E}_{n} \cdot \text{{PUE}}_{n}\ \ \  [\text{kWh}]
\end{equation}

\noindent
where $\cd{E}_n$ is the energy consumption caused by $s$ at $n$ (in kWh) and $\text{PUE}_n$ is the PUE of $n$. As aforementioned, $\cd{E}_{n}$ is a (non-linear) function of the current node load.
For each node \cd{N} involved in placement \cd{P}, \cd{hardwareFootprint/4} (line 9, 12--17) exploits \cd{hardwareEnergy/4} (line 14, 18--23) to first retrieve the node load \cd{OldL} before placing the services in placement \cd{P}, and the associated energy consumption \cd{OldE} (line 20). Then, it retrieves the node load \cd{NewL} after placing the services as per \cd{P} (line 21), and computes the associated energy consumption \cd{NewE} (line 22). The difference between \cd{NewE} and \cd{OldE}, multiplied by the PUE of \cd{N}, estimates the \cd{Energy} consumption of \cd{P} on node \cd{N} as per Eq.~(\ref{eq:energy}) (line 23).

Based on this, \greenbrain also estimates the associated carbon emissions. To this end, extending \cite{cotwo}, we consider the case in which multiple energy sources are combined at node $n$ -- each with an associated mix percentage $p_1, \dots, p_k$ such that $\sum_i p_i = 1$ -- producing $\mu_1, \dots, \mu_k$ emissions, respectively. 
Predicate \cd{hardwareEmissions/3} (line 15, 24--27) recursively scans the energy mix declared for node \cd{N} and computes carbon emissions as

\vspace{-3mm}
\begin{equation}\label{eq:immissions}
I_{s} = \cd{E}_{s} \cdot \sum_i p_i \mu_i\ \ \  [\text{\kgco}]
\end{equation}


\vspace{-2mm}
Finally, following the approach of \cite{microsoft}, \cd{networkFootprint/3} (lines 10, 28--31) estimates the carbon emissions to transmit traffic flows allocated by \cd{P}. Transmitting 1MB of data over the Internet requires around $0.00008$ kWh (\cd{kWhPerMB/1}) \cite{gci2} and the average \textit{global carbon intensity} (\cd{averageGCI/1}) of electricity is of 475 gCO2/kWh~\cite{gci}. Then, the network energy consumption $\cd{E}_N$ and carbon emissions $I_{N}$ for transmitting $M$ MB for one hour can be estimated as

\vspace{-3mm}
\begin{equation}\label{eq:bwen}
\cd{E}_N = 450 \cdot 0.00008 \cdot M\  [\text{kWh}]
\quad \text{ and } \quad
I_{N} = 0.475 \cdot \cd{E}_N \ [\text{\kgco}]
\end{equation}

\vspace{-2mm}
\noindent also considering that 1 Mbit/s = 450 MB/h. Eq.s (\ref{eq:bwen}) are computed at lines 30 and 31 of the code of Fig. \ref{fig:gFogBrain}, respectively.

\begin{figure}[!ht]
    \centering
\begin{Verbatim}[firstnumber=1,fontfamily=zi4, numbers=left, numbersep=5pt, fontsize=\footnotesize, frame=single, framesep=1mm, framerule=0.1pt, rulecolor=\color{gray}, tabsize=2,numberblanklines=false]
placements(A,Placements) :-
    findall((C,Cost,E,P), (gFogBrain(A,P,E,C), hourlyCost(P,Cost)), Ps),
    sort(Ps,Placements).

gFogBrain(A,P,Energy,Carbon) :- 
    application(A,Services), placement(Services,P),
    allocatedResources(P,Alloc), footprint(P,Alloc,Energy,Carbon).

footprint(P,(AllocHW,AllocBW),Energy,Carbon) :-
    deploymentNodes(P,Nodes), 
    hardwareFootprint(Nodes,AllocHW,HWEnergy,HWCarbon),
    networkFootprint(AllocBW,BWEnergy,BWCarbon),
    Energy is HWEnergy + BWEnergy, Carbon is HWCarbon + BWCarbon.

hardwareFootprint([(N,HW)|Ns],AllocHW,Energy,Carbon) :-
    hardwareFootprint(Ns,AllocHW,EnergyNs,CarbonNs),
    hardwareEnergy(N,HW,AllocHW,EnergyN), 
    energySourceMix(N,Sources), hardwareEmissions(Sources,EnergyN,CarbonN),
    Energy is EnergyN + EnergyNs, Carbon is CarbonN + CarbonNs.
hardwareFootprint([],_,0,0).

hardwareEnergy(N,HW,AllocHW,Energy):-
    totHW(N,TotHW), pue(N,PUE), 
    OldL is 100 * (TotHW - HW) / TotHW, energyProfile(N,OldL,OldE),
    findall(H,member((N,H),AllocHW),HWs), sum_list(HWs,PHW), 
    NewL is 100 * (TotHW - HW + PHW) / TotHW, energyProfile(N,NewL,NewE),
    Energy is (NewE - OldE) * PUE. 

hardwareEmissions([(P,S)|Srcs],Energy,Carbon) :-
    hardwareEmissions(Srcs,Energy,CarbSrcs),
    emissions(S,MU), CarbS is P * MU * Energy, Carbon is CarbS + CarbSrcs.
hardwareEmissions([],_,0).

networkFootprint(AllocBW,BWEnergy,BWCarbon) :-
    findall(BW, member((_,_,BW),AllocBW), Flows), sum_list(Flows,TotBW),
    kWhPerMB(K), BWEnergy is 450 * K * TotBW, 
    averageGCI(A), BWCarbon is A * BWEnergy.
\end{Verbatim}
\vspace{-6mm}
    \caption{Main predicates of \greenbrain.}
    \label{fig:gFogBrain}
\end{figure}

\example By querying \cd{placements(lightsApp,Placements)} over the inputs of Examples \ref{ex:app} and \ref{ex:infra}, we obtain the two eligible placements for application \textsf{LightsApp} listed in Table \ref{tab:results}, along with their estimated hourly carbon emissions, energy consumption, and cost.
Based on those and on business considerations, application operators can then informedly decide whether to enact $P_1$ or $P_2$. While $P_1$ saves more than $9\%$ \co emissions compared to $P_2$, and consumes $5\%$ less energy, it incurs in an $11\%$ cost increase (i.e. +0.004 \euro/h $\simeq$ +3 \euro /month). It is also possible to exploit \greenbrain to perform \textit{what-if} analyses and to possibly evaluate \textit{greener} infrastructure operators, thus improving on target metrics.
\hfill \qed

\vspace{-6mm}
\begin{table}[h]
    \caption{Example placement results.}
    \label{tab:results}
    \vspace{-2mm}
\centering
{\footnotesize
\begin{tabular}{|c|c|c|c|c|}
\hline
\textbf{Id} & \textbf{Placement}                                                                                         & \textbf{Emissions} & \textbf{Cost} & \textbf{Energy Cons.} \\ \hline
$P_{1}$     & \begin{tabular}[c]{@{}c@{}}\cd{on(lightsDriver, edgenode)}, \\ \cd{on(mlOptimiser, privateCloud)}\end{tabular}    & 0.29 \kgco              & 0.0356 \euro/h & 0.60 kWh                    \\ \hline
$P_{2}$     & \begin{tabular}[c]{@{}c@{}}\cd{on(lightsDriver, accesspoint)}, \\ \cd{on(mlOptimiser, privateCloud)}\end{tabular} & 0.32 \kgco              & 0.0316 \euro/h & 0.63 kWh                    \\ \hline
\end{tabular}}
\end{table}

%% file: src/relatedwork.tex
\section{Related Work}
\label{sec:related}
\vspace{-3mm}
Much work targeted the problem of placing multi-service applications in Cloud-IoT computing scenarios, e.g. as surveyed in~\cite{brogi2019place,applicationmanagementfogsurvey}. Only some works featured some aspects of energy-awareness but did not consider carbon footprint or relied on simple linear models for energy consumption (e.g.~\cite{barcelo2016iot,kopras2020latency,souza2016towards,yu2018green}).
To the best of our knowledge,  \cite{aldossary2021towards} is the first work including carbon emissions in the trade-off analyses to determine optimal Cloud-IoT application placements, via mixed integer linear programming. A limitation of \cite{aldossary2021towards} resides in the fact that it only considers linear energy consumption for infrastructure nodes. On the contrary, energy consumption is usually a non-linear function of a computational node load~\cite{microsoft,xiao2021electric}. Last, \cite{aldossary2021towards} does not consider the possibility to estimate energy consumption based on combined sources, do not account for operational costs estimates, and require full knowledge of the physical network topology and employed routing algorithms, which is not always available in real scenarios.

Focussing on declarative approaches, Casadei et al. \cite{casadei19,casadei21} devised a declarative approach to service coordination based on \textit{aggregate computing}, managing opportunistic resources via a hybrid centralised/decentralised solution by relying on a self-organising peer-to-peer architecture to handle churn and mobility. We have exploited logic programming to assess the security and trust levels of application placements \cite{secfog2019}, and to determine them \cite{fogbrain} also in Osmotic computing settings \cite{DBLP:conf/caise/0002B21}. Finally, we very recently proposed a fully decentralised solution to write and enforce QoS-aware application management policies written in Prolog~\cite{mario20}. None of those declarative solutions, however, considers energy consumption nor carbon emissions, as \greenbrain does.

%% file: src/conclusions.tex
\vspace{-2mm}
\section{Concluding Remarks}
\label{sec:conclusions}

\vspace{-2mm}
In this article, we have presented a declarative methodology and its prototype, \greenbrain, to determine eligible multiservice application placements and to estimate their carbon emissions, energy consumption and operational costs. The prototype allows application operators to determine application placements that satisfy all software, hardware, IoT, and network QoS constraints, and to informedly decide on the best trade-off placement considering its estimated impact on the environment and its deployment operational costs, which oftentimes represent contrasting objectives to optimise at once.

Future work includes improving \greenbrain with \textit{continuous reasoning} (as in~\cite{fogbrain}) and assessing it via simulation or over Cloud-IoT testbeds, also employing different formulas to estimate energy consumption and carbon emissions (including an amortised estimate of the deployment hardware \textit{embedded carbon}).  

%% file: src/appendix.tex
\section*{Appendix A -- Code of \cd{placement/2}}

For the sake of the Reviewers only, we report hereinafter the code of \cd{placement/4} from \cite{fogbrain}, exploited by predicate \cd{gFogBrain/4} (Fig. \ref{fig:gFogBrain}).

\begin{figure}[!h]
    \centering
\begin{Verbatim}[firstnumber=1,fontfamily=zi4, numbers=left, numbersep=5pt, fontsize=\scriptsize, frame=single, framesep=1mm, framerule=0.1pt, rulecolor=\color{gray}, tabsize=4,numberblanklines=false]
placement(Services,P) :- placement(Services, [], ([],[]), P).

placement([S|Ss],P,(AllocHW,AllocBW),Placement) :-
    nodeOk(S,N,P,AllocHW), linksOk(S,N,P,AllocBW), 
    placement(Ss,[on(S,N)|P],(AllocHW,AllocBW),Placement).
placement([],P,_,P).

nodeOk(S,N,P,AllocHW) :-
    service(S,SWReqs,HWReqs,IoTReqs),
    node(N,SWCaps,HWCaps,IoTCaps),
    swReqsOk(SWReqs,SWCaps),
    thingReqsOk(IoTReqs,IoTCaps),
    hwOk(N,HWCaps,HWReqs,P,AllocHW).

swReqsOk(SWReqs, SWCaps) :- subset(SWReqs, SWCaps).

thingReqsOk(TReqs, TCaps) :- subset(TReqs, TCaps).

hwOk(N,HWCaps,HWReqs,P,AllocHW) :-
    findall(HW,member((N,HW),AllocHW),HWs), 
    sum_list(HWs, CurrAllocHW),
    findall(HW, (member(on(S1,N),P), service(S1,_,HW,_)), OkHWs), 
    sum_list(OkHWs, NewAllocHW),  
    hwTh(T), HWCaps >= HWReqs + T - CurrAllocHW + NewAllocHW.

linksOk(S,N,P,AllocBW) :-
    findall((N1N2,ReqLat), distinct(relevant(S,N,P,N1N2,ReqLat)), N2Ns),
    latencyOk(N2Ns),
    findall(N1N2, distinct(member((N1N2,ReqLat),N2Ns)), N1N2s), 
    bwOk(N1N2s, AllocBW, [on(S,N)|P]). 

latencyOk([((N1,N2),ReqLat)|N2Ns]) :- 
    link(N1,N2,FeatLat,_), FeatLat =< ReqLat, latencyOk(N2Ns).
latencyOk([]).

bwOk([(N1,N2)|N2Ns],AllocBW,P) :-
    link(N1,N2,_,FeatBW),
    findall(BW, member((N1,N2,BW),AllocBW), BWs), 
    sum_list(BWs, CurrAllocBW), 
    findall(BW, s2sOnN1N2((N1,N2), P, BW), OkBWs), 
    sum_list(OkBWs, OkAllocBw), 
    bwTh(T), FeatBW  >=  OkAllocBw - CurrAllocBW + T, 
    bwOk(N2Ns,AllocBW,P).
bwOk([],_,_).

relevant(S,N,P,(N,N2),L) :- s2s(S,S2,L,_), member(on(S2,N2),P), dif(N,N2).
relevant(S,N,P,(N1,N),L) :- s2s(S1,S,L,_), member(on(S1,N1),P), dif(N1,N).

s2sOnN1N2((N1,N2),P,B) :- 
    s2s(S3,S4,_,B), member(on(S3,N1),P), member(on(S4,N2),P).

allocatedResources(P,(AllocHW,AllocBW)) :- 
    findall((N,HW), (member(on(S,N),P), service(S,_,HW,_)), AllocHW),
    findall((N1,N2,BW), n2n(P,N1,N2,BW), AllocBW).
n2n(P,N1,N2,ReqBW) :- 
    s2s(S1,S2,_,ReqBW), member(on(S1,N1),P), member(on(S2,N2),P), dif(N1,N2).
\end{Verbatim}
\vspace{-4mm}
    \caption{Declarative placement strategy of \cite{fogbrain}.}
    \label{fig:gFogBrain}
\end{figure}